\documentstyle[preprint,prl,aps]{revtex}
\begin{document}
\tightenlines
\draft

\title{Mixed state on a sparsely encoded associative memory model}

\author{Tomoyuki KIMOTO}
\address{
Department of Electrical Engineering,Oita National College of Technology
1666,Maki,Oita-shi,Oita 870-0152,Japan
}

\author{Masato OKADA}
\address{
Japan Science and Technology Corporation
2-2,Hikaridai,Seika-cho,Soraku-gun,Kyoto 619-0288,Japan
}

\date{\today}
\maketitle

\begin{abstract}
In the present paper,
we analyze symmetric mixed states
corresponding to the so-called {\it concept formation}
on a sparsely encoded associative
memory model with $0-1$ neurons.
Three types of mixed states,
OR, AND and a majority decision mixed state are described
as typical examples.
Each element of the OR mixed state is composed
of corresponding memory pattern elements
by means of the OR-operation.
The other two types are similarly defined.
By analyzing their stabilities through the SCSNA and the computer simulation,
we found that the storage capacity of the OR mixed state diverges
in the sparse limit, but that the other states do not diverge.
In addition, we found that the optimal threshold values,
which maximize the storage capacity,
for the memory pattern and the OR mixed state
coincide with each other in the spare limit.
Thus, we conclude that the OR mixed state is 
a reasonable representative of the mixed state in the spare limit.
Finally, the paper examines the relationship between our results
and recently reported physiological findings
regarding face-responsive neurons in the inferior temporal cortex.
\end{abstract}

\narrowtext

\section{Introduction}
\label{sec.intro}
In this paper, we analyze symmetric mixed states
(hereafter ``mixed state'') 
on a sparsely encoded associative memory model.
When an associative memory model is made to store memory patterns
as a result of correlation learning, 
a pattern, which is a nonlinear superposition of the stored memory patterns
using a neuron output function,
automatically becomes the equilibrium state of the model. 
This is called the mixed state.
It is not appropriate to think that this mixed state is 
a side-effect and/or that is unnecessary for information processing.
Amari has discussed a ``concept formation''
using the stability of mixed state \cite{Amari1977}. 
The correlated attractor proposed in \cite{Griniasty1993,Amit1994}, 
which is a model of the Miyashita attractor \cite{Miyashita1988b}, 
is considered to be a (not symmetric) mixed state in a broad sense. 
Recently, Parga and Rolls have used the mixed state
as a mechanism of invariant recognition
under a coordinate transformation \cite{Parga1998}.
On the other hand, 
the sparse coding scheme is believed to be used in the brain 
according to some physiological findings \cite{Miyashita1988b}
and the theoretical viewpoints 
\cite{Tsodyks1988,Buhmann1989,Amari1989,Perez-Vicente1989,Okada1996}.
Thus, the properties of the mixed states on the sparsely
encoded associative memory model should be discussed. 
This is the main purpose of the present paper.

$\mbox{sgn}(\sum_{\mu=1}^s \xi^\mu_i+h)$ is the mixed state
in a model storing memory pattern $\mbox{\boldmath $\xi$}^\mu$,
with the output function $\mbox{sgn}(\cdot)$ and threshold value $h$.
Therefore $s$ types of mixed state can be composed of 
$s$ memory patterns by changing the threshold value.
$\mbox{sgn}(\sum_{\mu=1}^s \xi^\mu_i)$ among the $s$ types is regarded 
as the mixed state 
in a model with the output function $\mbox{sgn}(\cdot)$
that is made to learn memory patterns 
of a $50\%$ firing rate \cite{Amit1985a}. 
In other words, 
a ``typical'' mixed state is a state of the majority decision 
for the firing or non-firing elements among the $s$ types of the mixed state.
A natural question then arises:
what kind of mixed state among the $s$ types should be considered 
typical in the sparse limit of $f \rightarrow 0$ ? 
($f$ is the firing rate of the memory pattern). 
In the sparsely encoded model, 
storage capacity strongly depends on the threshold value,
so that an appropriate setting of the threshold value 
makes the storage capacity 
diverge at $1/|f \log f|$  in the sparse limit 
\cite{Tsodyks1988,Buhmann1989,Amari1989,Perez-Vicente1989,Okada1996}.
Thus, in this study, 
``typical'' mixed state is defined as the mixed state 
with similar properties to the memory pattern
for the storage capacity and the threshold value.
The purpose of this paper is to discuss which mixed state 
in the sparse limit should be regarded as a typical mixed state. 

First, 
three out of $s$ types of the mixed state are considered: 
the OR mixed state, the majority decision mixed state, 
and the AND mixed state. 
Each element of the OR mixed state is taken from the OR operation 
in the Boolean sense
for each element of memory patterns composing of the mixed state. 
The other two types are similarly defined.
The examination of the stability of 
each mixed state by the self-consistent signal-to-noise analysis (SCSNA) 
proposed by Shiino and Fukai \cite{Shiino1992} and the computer simulation 
shows that the storage capacity of the OR mixed state diverges
as $1/|f\log f|$, 
but the storage capacities of the other two types do not diverge. 
This can be explained as follows: 
the overlap of the majority decision mixed state
with the stored memory pattern
and the overlap of the AND mixed state with the stored memory pattern
converge at $0$ in the sparse limit, 
whereas the overlap of the OR mixed state with one converges at $1$. 
Likewise, while the optimum threshold value 
that maximizes the storage capacity of the memory pattern or the OR
mixed state settles at the same value, 
the optimum threshold value of the majority decision mixed state and the 
AND mixed state does not agree with the optimum threshold value 
of the memory pattern
in the sparse limit. 
Therefore, 
the important factors 
to decide the storage capacity and the optimum threshold value
in the sparse limit are asymptotes for the overlaps
of the $s$ types of the mixed state with the memory patterns.
Besides three typical mixed states, 
we also evaluate the overlaps of the $s$ types of the mixed states 
to show that all the overlaps 
of the $s-1$ kinds of mixed states except for the OR mixed states converge 
to $0$ in the sparse limit. 
Thus, 
we conclude that a consideration 
of the OR mixed state as a typical mixed state 
in the sparse coding is appropriate. 

At the end of this paper, 
we discuss 
the recently reported physiological findings of Sugase et al. \cite{Sugase1998}
on the face-responsive neurons in the inferior temporal cortex. 
They have reported that some face-responsive neurons in the IT cortex
code rough classification information in the first half of the firing 
and then code detailed classification information
 in the latter half of the firing. 
Their findings suggest that
the OR mixed state emerges 
in the first half of the neuron's dynamics,
and then the network state converges in the memory state.
We will infer a relationship between the physiological findings
and our present results.

\section{Model}
Shown below are the equilibrium properties of a recurrent neural network 
composed of $N$ neurons with an output function $\Theta(\cdot)$,
\begin{eqnarray}
 x_i &=& \Theta (u_i), \\
 u_i &=& \sum_{j \ne i}^N J_{ij} x_j + h,
  \label{eq:equilibrium}
\end{eqnarray}
\begin{equation}
	\Theta(u)  = \cases{
	      1 & u $\ge$ 0 \cr
        0 & u  $<$  0 \cr
      } ,
\label{eq.theta}
\end{equation}
where $x_i$ denotes an output of the $i$-th neuron in the equilibrium state, 
and $J_{ij} $ denotes a synaptic weight from the $j$-th neuron 
to the $i$-th neuron.
The threshold value $h$ of the neuron is assumed to not depend 
on the serial number $i$ of a neuron. 
Its concrete value is described later. 
Each element $\eta^\mu_i$ of the $\mu$-th memory pattern
$\mbox{\boldmath $\eta$}^\mu$,
which is stored in the present model, 
is independently generated
by the probability, 
\begin{equation}
\mbox{Prob}[\eta^\mu_i=1] = 1-\mbox{Prob}[\eta^\mu_i=0] = f,
  \label{eq.pattern}
\end{equation}
where $\mbox{E}[\eta^{\mu}_i]= f$, and $f$ stands for the firing rate of
memory pattern $\mbox{\boldmath $\eta$} ^\mu$ .
The pattern with a small firing rate $f$ is called
a {\it sparse} pattern,
and the use of a sparse pattern for the memory pattern is called a sparse
encoding.
The synaptic weight $J_{ij}$ is decided by the covariance learning method,
\begin{equation}
 J_{ij}=\frac {1}{Nf(1-f)} \sum_{\mu=1}^{\alpha N}
  (\eta_i^\mu-f) (\eta_j^\mu-f),
  \label{eq:covariance}
\end{equation}
where $\alpha N$ is a number of the stored memory patterns, 
and $\alpha$ is defined as the loading rate.

Let us define the mixed state
$\mbox{\boldmath{$\gamma$}} ^{(s,k)}$ composed of $s$ memory patterns.
The $i$-th element of the mixed state
$\mbox{\boldmath{$\gamma$}}^{(s,k)}$ is set to $'1'$ 
if the number of the firing state ($'1'$) is $k$ or more 
in the $i$-th elements of the $s$ memory patterns
that compose the mixed state. 
Otherwise, it becomes $'0'$.
The $s$ types of the mixed states exist 
according to this definition
because $1 \leq k \leq s$.
In particular, among $s$ types of mixed state, 
mixed state $\mbox{\boldmath{$\gamma$}}^{(s,1)}$
 is considered to be the OR mixed state,
where its $i$-th element is given by 
the OR operation through the $i$-th elements
of $s$ memory patterns.
Following the definition of the OR mixed state,
$\mbox{\boldmath{$\gamma$}}^{(s,s)}$ corresponds to 
the AND mixed state where its $i$-th element is obtained by the AND operation.
Moreover, $\mbox{\boldmath{$\gamma$}} ^{(s,s/2)} $ is 
the majority decision mixed state that regards the one 
with more numbers of 0 or 1 for the $i$-th elements of $s$ memory patterns.

The threshold value $h$ in Eq.(\ref{eq:equilibrium}) is 
decided as follows.
The threshold value $h$ should be appropriately chosen
in the sparsely encoded model
\cite{Tsodyks1988,Buhmann1989,Amari1989,Perez-Vicente1989,Okada1996}.
In this paper, 
the threshold value is calculated by using
the mean firing rate of the retrieval pattern.
The threshold value obtained by this method 
approximately coincides with the optimum threshold value 
by which the storage capacity is maximized \cite{Okada1996}.  
Therefore, we employ a method with a constant firing rate 
as an approximate method of obtaining the optimum threshold value
in the present paper. 
Since the mean firing rate of the memory pattern is $f$, 
when the memory pattern is retrieved, 
threshold value $h$ is decided from the following equation,
\begin{equation} 
 f = \frac{1}{N} \sum_{i=1}^N
  \Theta \left(\sum_{j \neq i}^N J_{ij} x_j + h \right).
\label{eq.threshold value}
\end{equation}
Moreover, 
the mean firing rate $f^{(s,k)}$ 
of mixed state $\mbox{\boldmath{$\gamma$}} ^{(s,k)}$ replaces 
$f$ in Eq. (\ref{eq.threshold value}) 
to recall the mixed state $\mbox{\boldmath{$\gamma$}} ^{(s,k)}$, 
and threshold value $h$ is decided,
\begin{equation} 
 f^{(s,k)} = \mbox{E}[\gamma^{(s,k)}_i]
  = \sum_{v=k}^{s} {}_s \mbox{C}_{v} f^v (1-f)^{s-v}.
  \label{eq.f_s_k}
\end{equation}
Here, $\mbox{C}$ is a number of the combination.
The overlap of the equilibrium state $\mbox{\boldmath{$x$}}$ 
with the $\mu$-th memory pattern $\mbox{\boldmath{$\eta$}}^\mu$ 
is defined,
\begin{equation}
  m_\mu = \frac{1}{Nf(1-f)}
  \sum_{i=1}^N (\eta_i^\mu-f) x_i.
  \label {eq:overlap}
\end{equation}
If the equilibrium state $\mbox{{\boldmath{$x$}}}$ is completely equal
to $\mbox{\boldmath{$\eta$}}^\mu$, then $m_\mu=1$.
The overlap $M^{(s,k)}$ of the equilibrium state {\boldmath{$x$}} 
with the mixed state $\mbox{\boldmath{$\gamma$}} ^{(s,k)}$ 
is defined in a manner similar to the Eq. (\ref{eq:overlap}),
\begin{equation}
  M^{(s,k)} = \frac{1}{N f^{(s,k)} (1- f^{(s,k)} )}
  \sum_{i=1}^N (\gamma^{(s,k)}_i - f^{(s,k)}) x_i.
  \label {eq:overlapM}
\end{equation}
If the equilibrium state $\mbox{{\boldmath{$x$}}}$ is 
$\mbox{\boldmath{$\gamma$}} ^{(s,k)}$, then $M^{(s,k)} =1$.

\section{Results}
\subsection{SCSNA analysis }
\label{sec.result1}
The SCSNA \cite{Shiino1992} and a computer simulation
are used to analyze
stabilities of three typical types of mixed state, 
the OR mixed state $\mbox{\boldmath{$\gamma$}} ^{(s,1)}$, 
the majority decision mixed state $\mbox{\boldmath{$\gamma$}} ^{(s,s/2)}$, 
and the AND mixed state $\mbox{\boldmath{$\gamma$}} ^{(s,s)} $. 
The order parameter equations of the SCSNA are derived 
as shown in the Appendix.

First, 
the OR mixed state $\mbox{\boldmath{$\gamma$}} ^{(s,1)} $is described.
Fig. \ref{fig:1} shows the loading rate $\alpha$ dependency 
of overlap $M^{(3,1)}$ on the OR mixed state 
$\mbox{\boldmath{$\gamma$}} ^{(3,1)}$ 
composed of three memory patterns ($s=3$).
The three lines indicate the results 
of the SCSNA and the data points and the error bars are the results 
of the computer simulation
for the mean firing rate of the memory pattern, $f=0.5, 0.2$ and $0.1$. 
In the computer simulation, the neuron number 
is set as $N \geq 10,000$, and the simulation is performed 11 times 
for each parameter. 
The data point shows the median, 
and both ends of the error bar show the $1/4$ deviation 
and the $3/4$ deviations, respectively.
The horizontal axis is the loading rate $\alpha$, 
and the vertical axis is the overlap $M^{(3,1)}$ 
of the OR mixed state $\mbox{\boldmath{$\gamma$}} ^{(3,1)}$ 
with the equilibrium state $\mbox{\boldmath{$x$}}$.
The results of the SCSNA and the computer simulation correspond
well with each other 
for the loading rate $\alpha$ dependency of overlap $M^{(3,1)}$ 
and the storage capacity $\alpha_C$, which is the loading rate 
when the equilibrium state becomes unstable.
The properties of the three types of the mixed states are examined 
using the SCSNA results. 
This is because the SCSNA results explain 
the computer simulation results fairly well,
as shown in Fig. \ref{fig:1}.
Fig. \ref{fig:2} shows the mean firing rate $f$ dependency of 
the storage capacity $\alpha_C$ on
$\mbox{\boldmath{$\gamma$}} ^{(s,1)} \; (s=3,5,7)$ 
of the OR mixed state.
The solid line shows 
the storage capacity of memory pattern ($s=1$).
The higher the mixed number $s$, 
the smaller the storage capacity is with the same mean firing rate $f$. 
However, the lower the mean firing rate $f$, 
the larger the storage capacity for all three states of $s$.
Then, we examine the asymptotes for $f \rightarrow 0$.
The asymptotes for the all values of $s$ shown 
in Fig. \ref{fig:3} are the same $1/|f \log f|$ 
as the case for the memory patterns.
In addition, 
the storage capacity of the OR mixed state 
$\mbox{\boldmath{$\gamma$}} ^{(s,1)}$ 
with the $s$ memory patterns with firing rate $f$ 
is the same as that of the memory pattern 
with firing rate $sf$ in the sparse limit. 
The reason is explained in the next section \S \ref{sec.result2}.
Fig. \ref{fig:4} shows that the mean firing rate $f$ dependency 
of threshold value $h_c$ at the loading rate $\alpha$ is set to the 
storage capacity, i.e., $\alpha=\alpha_C$.
The figure shows the threshold value of the memory pattern 
with a solid line for comparison.
In the limit of $f \rightarrow 0$, 
threshold value $h_c$ converges 
to the same value without depending 
on the $s$ that includes the memory pattern $s=1$.
The reason is also explained in the next section \S \ref{sec.result2}.
These findings imply that both of the memory patterns
$\mbox{\boldmath{$\eta$}} ^\mu$
and the OR mixed state $\mbox{\boldmath{$\gamma$}} ^{(s,1)}$ 
easily coexist as stable in the sparse limit $f \rightarrow 0$.

Next, we examine the majority decision mixed state
$\mbox{\boldmath{$\gamma$}}^{(s,s/2)}$ 
and the AND mixed state $\mbox{\boldmath{$\gamma$}} ^{(s,s)}$.
Fig. \ref{fig:5} and \ref{fig:6} show the storage capacity 
for the majority decision mixed state
$\mbox{\boldmath{$\gamma$}}^{(s,s/2)}, \; (s=3,5,7)$ 
and the AND mixed state
$\mbox{\boldmath{$\gamma$}}^{(s,s)}, \; (s=3,5,7)$, respectively.
Fig. \ref{fig:5} shows that the storage capacity gradually increases 
in $s=3$ as the mean firing rate $f$ decreases, 
and the storage capacity decreases again 
without diverging in the sparse limit.
It is about $\alpha_c=0.065$ even at the maximum.
If the number $s$ of memory patterns is increased to $s=5, 7$, 
the maximum storage capacity is reduced further, 
and it goes toward 0 in the sparse limit.
Fig. \ref{fig:6} shows that the storage capacity $\alpha_c$ 
of the AND mixed state $\mbox{\boldmath{$\gamma$}}^{(s,s)}$ decreases 
as the mean firing rate becomes small in each number $s$ of the mixed
state, and it goes toward 0 in the sparse limit.
These results show that storage capacity $\alpha_C$ does not diverge 
in the sparse limit except in the memory pattern 
and the OR mixed state $\mbox{\boldmath{$\gamma$}}^{(s,1)}$.
The reason is qualitatively explained in \S \ref{sec.result2}.
Fig. \ref{fig:7} and \ref{fig:8} show 
the mean firing rate $f$ dependency of threshold value $h_c$ 
at the storage capacity $\alpha_c$
for the majority decision mixed state 
$\mbox{\boldmath{$\gamma$}}^{(s,s/2)}, \; (s=3,5,7)$ 
and the AND mixed state
$\mbox{\boldmath{$\gamma$}}^{(s,s)}, \; (s=3,5,7)$.
The figure shows the threshold value of the memory pattern 
with a solid line for comparison.
The threshold values of these types of mixed state 
do not correspond to the threshold value of the memory pattern 
in the sparse limit, 
while that of the OR mixed state $\mbox{\boldmath{$\gamma$}}^{(s,1)}$
is the same as that of the memory pattern.
This means that the memory pattern and these types of mixed state 
cannot coexist 
in the sparse limit as stable.
On the other hand, the threshold value of the majority decision mixed state 
$\mbox{\boldmath{$\gamma$}}^{(s,s/2)}$ crosses the threshold value 
of the memory pattern at the mean firing rate of $f=0.5$.
As mentioned chapter \S \ref{sec.intro}, this implies that 
the majority decision mixed state $\mbox{\boldmath{$\gamma$}}^{(s,s/2)}$ 
can be considered as the typical mixed state in $f=0.5$
\cite{Amit1985a}.

\subsection{Qualitative evaluation of mixed state by the naive S/N
analysis}
\label{sec.result2}
In this section, 
we explain the analytical results of the SCSNA
by using the naive S/N analysis
\cite{Perez-Vicente1989,Okada1996}. 
First, 
the reason why only the storage capacity of the OR mixed state 
$\mbox{\boldmath{$\gamma$}} ^{(s,1)}$ diverges in the sparse limit is
discussed.
Assuming that output $x_i$ of each neuron is equal to the mixed state 
$\gamma^{(s,k)}_ i$, 
the internal potential $u_i$ is rewritten with 
the synaptic weight $J_{ij}$ 
of Eq. (\ref{eq:covariance}).
Without loss of generality,
the mixed state $\mbox{\boldmath{$\gamma$}}^{(s,k)}$ 
is composed of $\mbox{\boldmath $\eta$}^\mu, \; (1 \le \mu \le s)$, 
because the synaptic weight $J_{ij}$ in Eq. (\ref{eq:covariance}) 
is invariant for the replacement of the order of $\mu$.
The $u_i$ in Eq. (\ref{eq:equilibrium}) 
is decomposed into $1 \le \mu \le s$ and other parts,
\begin{eqnarray}
 u_i &=& \sum_{j \neq i}^N J_{ij} \gamma^{(s,k)}_j + h \nonumber \\
 &=& \sum_{\mu=1}^{s} (\eta^\mu_i-f) m^{(s,k)}_{\mu}  + h + \bar{z},
    \label{eq.S/N} \\
 m^{(s,k)}_\mu &=& \frac{1}{Nf(1-f)} \sum_{i=1}^N
  (\eta^\mu_i -f) \gamma^{(s,k)}_i, 
\end{eqnarray}
\begin{eqnarray}
    \bar{z_i} &=& \frac{1}{Nf(1-f)} \sum_{\mu=s+1}^{\alpha N}
      \sum_{j \ne i}^{N} (\eta^\mu_i-f)(\eta^\mu_j-f) \gamma^{(s,k)}_j .
      \nonumber \\
     & &
      \label{eq.naive_noise}
\end{eqnarray}
The first term of Eq. (\ref{eq.S/N}), which is composed of 
finite number of overlap $m^{(s,k)}_{\mu}$,
is a signal term to retrieve.
The second term is the threshold value, 
and $\bar{z}_i$ the third term is a cross-talk noise,
which prevents the state $\mbox{\boldmath $\gamma$}^{(s,k)}$ from being stable.
In this case, $\bar{z_i}$ obeys the normal distribution 
$N(0, \alpha f^{(s,k)})$ in the limit of $N \rightarrow \infty$.
Let us evaluate $m^{(s,k)}_{\mu}$ in the sparse limit.
However, we examine $k \ge 1$ 
because all elements of the mixed state with $k=0$ become $1$.
The overlap $m^{(s,k)}_{\mu}$ of
the mixed state $\mbox{\boldmath{$\gamma$}}^{(s,k)}$ 
with the memory pattern $\mbox{\boldmath $\eta$}^\mu$ 
can be calculated 
by using the expectation value 
in memory pattern $\mbox{\boldmath $\eta$}^\mu$ 
generated with the probability in Eq. (\ref{eq.pattern}),
\begin{eqnarray}
   m^{(s,k)}_{\mu} &=& \frac{1}{f(1-f)} 
    \mbox{E} [ (\eta_i^\mu-f) \gamma^{(s,k)}_i ]
     \label{S/N:Em} \\
           &=& \sum_{v=0}^{s-1} 
	    {}_{s-1} \mbox{C}_{v} f^v (1-f)^{s-v-1} \nonumber \\
           & & \times \left[ \Theta(1+v-k)-\Theta(v-k) \right] .
     \label{S/N:m}
\end{eqnarray}
Here $\mbox{C}$ is a number of the combination.
Note that $\Theta(0) =1$ as in Eq. (\ref{eq.theta}).
In the sparse limit of $f \rightarrow 0$, 
$f^v$ in Eq. (\ref{S/N:m}) becomes 1 in $v=0$ 
and 0 in $v \ge 1$.
Consequently, the summation for $v$ of Eq. (\ref{S/N:m}) 
has to be taken into account only when $v=0$.
Thus, $(1-f)^{s-v-1} \rightarrow 1$ for $f \rightarrow 0$,
and ${}_{s-1} \mbox{C}_{v}=1$ for $v=0$.
$\Theta (0-k) = 0$ as a result of taking $k \ge 1$ into consideration.
Therefore, the Eq. (\ref{S/N:m}) becomes the following in the sparse limit,
\begin{eqnarray}
   m^{(s,k)}_{\mu} = \Theta(1-k) .
     \label{S/N:m_f0}
\end{eqnarray}
This expression explains that $m^{(s,k)}_{\mu}=1$ only when $k=1$, 
and $m^{(s,k)}_{\mu}=0$ in $k\ge2$ regardless of the number of $s$ 
in the sparse limit.
Fig. \ref{fig:9} shows the mean firing rate $f$ dependency 
of the overlap of $m^{(5,k)}_{\mu}$.
This figure also shows that only when $k=1$, 
$m^{(s,k)}_{\mu} \rightarrow 1$ in the sparse limit, 
and $m^{(s,k)}_{\mu} \rightarrow 0$ in $k\ge2$.

If we use Eq. (\ref{eq.f_s_k}), 
the variance $\alpha f^{(s,k)}$ of cross-talk noise $\bar{z_i}$ 
in Eq. (\ref{eq.naive_noise}) becomes,
\begin{equation}
 \alpha f^{(s,k)} =
  \alpha \sum_{v=k}^{s} {}_s \mbox{C}_{v} f^v (1-f)^{s-v}.
  \label{eq.naive_noise_variance}
\end{equation}
The mean firing rate $f^{(s,1)}$ of 
the $\mbox{\boldmath{$\gamma$}}^{(s,1)}$ converges on
$f^{(s,1)} \rightarrow sf$ in the sparse limit.
Therefore, in the sparse limit, 
the variance of cross-talk noise 
when the memory pattern with the $sf$ mean firing rate is recalled,
is $\alpha s f$ .
That is the case of the mixed state composed a memory pattern 
with $f$ mean firing rate is $\alpha s f$.
Thus, they are the same as each other in the sparse limit.
This is the reason why the storage capacity of the OR mixed state 
$\mbox{\boldmath{$\gamma$}}^{(s,1)}$, 
which is composed of $s$ memory patterns with the mean firing rate $f$, 
is the same as that 
of the memory pattern with the mean firing rate $sf$.
In the same way, when the mixed state $\mbox{\boldmath $\gamma$} ^{(s,k)}$, 
which is $k \geq 2$, 
is evaluated on the variance of the cross-talk noise 
in Eq. (\ref{eq.naive_noise_variance}), it becomes 0 in the sparse limit. 
Thus, both $m^{(s,k)}_\mu$ constituting the signal term 
and variance for the cross-talk noise term become 0 in the sparse limit. 
When the S/N ratio at that time is calculated for $k \geq 2$, 
it converges to $0$.
Therefore, the study showed that the storage capacity 
did not become 0 only in the OR mixed state
$\mbox{\boldmath{$\gamma$}}^{(s,1)}$ of $k=1$ in the sparse limit.

Here we use the naive S/N analysis to discuss 
why the threshold value of only OR mixed state 
$\mbox{\boldmath{$\gamma$}} ^{(s,1)}$ converges at the value excluding 0,
while the threshold value converges at $0$ for $k\ge2$
in the sparse limit. 
Threshold value $h$ is a boundary value
where the neuron state $x_i$ with the internal potential $u_i$
is assigned to either $1$ or $0$ such as $x_i = \Theta(u_i + h)$.
Thus, if the absolute value of the signal term approaches $0$, 
the threshold value,
which is the boundary value ,
goes toward $0$.
This is why the threshold value converges at $0$ 
on the mixed state with $k \ge 2$
where the absolute value of the signal term goes toward $0$ 
in the sparse limit.
On the other hand, $m^{(s,1)}_\mu \rightarrow 1$ on the OR mixed state 
$\mbox{\boldmath{$\gamma$}}^{(s,1)}$,
and the overlap $m^{(1,1)}_\mu$ on the memory pattern is
always $m^{(1,1)}_\mu=1$.
Therefore, the threshold value that stabilizes 
the OR mixed state $\mbox{\boldmath{$\gamma$}}^{(s,1)}$ 
and the threshold value that stabilizes the memory pattern
coincide with each other at $h=-0.5$ in the naive S/N analysis. 
The threshold value obtained from the SCSNA 
as shown in Fig. \ref{fig:4} becomes $h_C=-0.7$, 
and it differs somewhat from the result $h=-0.5$.
However, this is the qualitative reason for the optimum threshold value 
of the OR mixed state $\mbox{\boldmath{$\gamma$}} ^{(s,1)}$
that corresponds to the threshold value of the memory pattern in the sparse
limit $f \rightarrow 0$.

\section{Summary and Discussion}
The various types of symmetric mixed states on 
the sparsely encoded associative memory model
composed of $0-1$ neurons were investigated using the SCSNA 
and the computer simulation.
The results showed that
the storage capacity of the OR mixed state diverges in manner 
similar to the memory pattern in the sparse limit with $1/|f \log f|$,
but that the storage capacities of the other types of the mixed states 
do not diverge. 
In order to clarify this reason, 
we evaluated the overlaps of $s$ types of the mixed states 
with the memory pattern. 
The overlap of the OR mixed state converged 
to $1$ in the sparse limit,
while the overlaps of all of the other $s-1$ types of the mixed states 
converged to $0$. 
Moreover, the evaluation of the variance of the cross-talk noise 
on the naive S/N analysis provided the above qualitative causes. 
The threshold value of the OR mixed state 
corresponded to the threshold value of the memory pattern in the sparse limit.
This was understood from an investigation of the overlap mentioned above. 
We conclude that the OR mixed state 
is the appropriate mixed state in the sparse encoding scheme. 

Finally, let us infer a relationship 
between the present theoretical results 
and recently reported physiological findings 
of Sugase et al. \cite{Sugase1998} for the face-responsive cell in
the IT cortex.
Sugase et al. performed a single unit recording 
in the IT cortex of two macaque monkeys 
by showing them various visual stimuli, such as 
monkey- and human-faces with various facial expressions 
and simple geometric shapes. 
They measured the temporal change of the information 
carried by the firing of neurons
for the classification of visual stimulus sets \cite{Kitazawa1998}.
They found that the initial transient firing correlated well with rough
categorizations (e.g. face vs. nonface stimuli), and the subsequent
sustained firing represented more detailed information. 
Their results suggest that
the neuron firing pattern is initially a superposition of patterns
representing different faces or expressions,
but it then converges to a single pattern
representing a specific face or expression.
To reinterpret this transient phenomenon 
by the "words" discussed in this paper, 
we might be able to say that the OR mixed state appears 
in the initial part of dynamics of neurons, 
and the network state finally converges on the memory pattern.
To our regret, such transient phenomenon did not occur in this model. 
Recently, Toya et al. and Okada et al. have discussed
an associative memory model with hierarchically correlated memory patterns 
\cite{Toya,OkadaNIPS}.
The hierarchically correlated patterns employed in \cite{Toya}
may capture the qualitative properties
of the visual stimuli used by Sugase et al.
Okada et al.  reported the following transient phenomena 
that were similar to the results of Sugase et al..
The network state approached a mixed state at the first stage
of the retrieval process.
After that it diverged away from the mixed state
and finally converged with the memory pattern
that was closest to the input pattern.
However, in their model,
the mixed state was not the OR mixed state 
but the majority decision mixed state.
It was thought to be the majority decision mixed state 
because the memory pattern was made to be 50$\%$ of the firing rate 
in \cite{Toya}. 
As mentioned before, 
the typical mixed state for the 50$\%$ firing rate  
is the majority decision mixed state. 
Considering our results and the results of \cite{Toya},
we conjecture that 
the associative memory model stores the sparsely encoded memory patterns 
with a hierarchical structure, 
and the state of the model is not directly drawn into any memory pattern, 
but drawn into the memory patterns after being drawn once 
into the OR mixed state. 
This prediction agrees with the behavior of IT cortex neurons 
suggested by Sugase et al.


\appendix
\section{}

The order parameter equation of the SCSNA is given as follows
\cite{Shiino1992},
\begin{equation}
 Y = F \left( \sum_{\mu=1}^{s} (\eta^{\mu}-f) m^\mu
 + h + \Gamma Y + \sigma z\right),
  \label{eq:Y}
\end{equation}
\begin{eqnarray}
 m^\nu &=& \int^\infty_{-\infty}
  Dz <\frac{\eta^{\nu}-f}{f(1-f)} Y>,
  \label{eq:m}
  \\
 q &=& \int^\infty_{-\infty}
  Dz \; <Y^2>,
  \label{eq:q}
  \\
 U &=& \frac 1 \sigma
  \int^\infty_{-\infty}
  Dz \; z <Y>,
  \label{eq:U}
  \\
 \sigma^2 &=& \frac{\alpha q}{(1-U)^2},
  \label{eq:sigma} 
  \\
 \Gamma &=& \frac{\alpha U}{1-U}.
  \label{eq:Gamma}
\end{eqnarray}
where,
\begin{equation}
 Dz = \frac {dz}{\sqrt {2 \pi}} \exp(-\frac{z^2}2).
  \label{eq:GaussFunc}
\end{equation}
Here $< \; >$ stands for the average over the elements 
of retrieved memory pattern $\eta^\mu \; (1 \leq \mu \leq s) $ 
that obeys Eq. (\ref{eq.pattern}).
Since the output function is $F(\cdot) = \Theta(\cdot)$, 
we apply the Maxwell rule \cite{Shiino1992,Okada1996}, 
and obtain the following expression,
\begin{equation}
Y(\eta,\alpha,z) = 
 \Theta \left( \sum_{\mu=1}^{s} (\eta^{\mu}-f) m_{\mu}
   + h + \frac{\Gamma}{2} + \sigma z\right).
  \label{eq:Y1}
\end{equation}
We also discuss the symmetric mixed state in this paper, 
where it is possible to put $m^\mu=m \; (1 \leq \mu \leq s) $. 
By substituting these conditions into Eq. (\ref{eq:Y}) 
to (\ref{eq:GaussFunc}), we obtain the following Eq. (\ref{eq.m}) 
to (\ref{eq.U}),
\begin{eqnarray}
  m &=& 
  \frac{1}{2} \sum_{v=0}^{s-1} {}_{s-1} \mbox{C}_{v} f^v (1-f)^{s-v-1} 
  	\nonumber \\
  & & \!\!\!\!\!\!\!\!\!\! \times \bigg[
	  \mbox{erf} (\frac{ m (v-sf+1) + h + \frac{\Gamma}{2}} {\sqrt{2}\sigma})
	    \nonumber \\
	& & \!\!\!\!\!\!\!\!\!\!\!\!
	- \mbox{erf} (\frac{ m (v-sf) + h + \frac{\Gamma}{2}} {\sqrt{2}\sigma}) 
  \bigg],
 \label{eq.m}
\\
 q &=& 
 \frac{1}{2} + \frac{1}{2} \sum_{v=0}^{s} {}_{s} \mbox{C}_{v}
    f^v (1-f)^{s-v} \nonumber \\
    & & \times \mbox{erf} (
 		\frac{ m (v-sf) + h + \frac{\Gamma}{2} }{\sqrt{2}\sigma} ),
 		\label {eq.q}
\\
 U &=& \frac{1}{\sqrt{2\pi}\sigma} 
  \sum_{v=0}^{s} {}_{s} \mbox{C}_{v}
  f^v (1-f)^{s-v} \nonumber \\
  & & \times \exp \big( -(
 		\frac{ m (v-sf) + h + \frac{\Gamma}{2} }{\sqrt{2}\sigma} )^2
 		\big).
 \label{eq.U}
\end{eqnarray}
Eq. (\ref{eq.threshold value}) decides the threshold value $h$ 
so that,
\begin{eqnarray}
 f &=& 
 \frac{1}{2} + \frac{1}{2} \sum_{v=0}^{s} {}_{s} \mbox{C}_{v} f^v (1-f)^{s-v}
 \nonumber \\
    & & \times \mbox{erf} (
 		\frac{ m (v-sf) + h + \frac{\Gamma}{2} }{\sqrt{2}\sigma} ).
\label{eq:threshold value}
\end{eqnarray}
For the mixed state $\mbox{\boldmath $\gamma$} ^{(s,k)} $, $f$ 
in the left term of Eq. (\ref{eq:threshold value}) 
only has to be changed into $f^{(s,k)}$.
Eq. (\ref{eq.f_s_k}) gives the mean firing rate $f^{(s,k)}$.
The overlap $M^{(s,k)}$ of an equilibrium state 
with the mixed state $\mbox{\boldmath $\gamma$} ^{(s,k)} $ 
in Eq. (\ref{eq:overlapM}) is shown as,
\begin{eqnarray}
 M^{(s,k)} &=& 
    \sum_{v=0}^{s} \frac{\Theta(v-k)-f^{(s,k)}}{2f^{(s,k)} (1-f^{(s,k)})} 
    {}_{s} \mbox{C}_{v} f^v (1-f)^{s-v} \nonumber \\
    & & \times \mbox{erf} ( 
    \frac{ m (v-sf) + h + \frac{\Gamma}{2} }{\sqrt{2}\sigma} ).
  \label {eq.M}
\end{eqnarray}



\begin{figure}
\caption{ 
\leftline{ Storage rate $\alpha$ dependency of overlap $M^{(3,1)}$ on OR mixed state $\mbox{\boldmath $\gamma$} ^{(3,1)} $of $s=3$ }
}
\label{fig:1}
\end{figure}

\begin{figure}
\caption{ 
\leftline{ Storage capacity $\alpha_C$ on OR mixed state $\mbox{\boldmath{$\gamma$}}^{(s,1)}$ }
}
\label{fig:2}
\end{figure}

\begin{figure}
\caption{ 
\leftline{Asymptotic characteristic of storage capacity $\alpha_C$ on OR mixed state $\mbox{\boldmath{$\gamma$}} ^{(s,1)}$. }
}
\label{fig:3}
\end{figure}

\begin{figure}
\caption{ 
\leftline{ Mean firing rate $f$ dependency of optimum threshold
 value $h_c$ on OR mixed state $\mbox{\boldmath{$\gamma$}}^{(s,1)}$ }
}
\label{fig:4}
\end{figure}

\begin{figure}
\caption{ 
\leftline{ Mean firing rate $f$ dependency of storage capacity 
on majority decision mixed state $\mbox{\boldmath{$\gamma$}} ^{(s,s/2)}$ }
}
\label{fig:5}
\end{figure}

\begin{figure}
\caption{ 
\leftline{ Mean firing rate $f$ dependency of storage capacity 
on AND mixed state $\mbox{\boldmath{$\gamma$}} ^{(s,s)}$ }
}
\label{fig:6}
\end{figure}

\begin{figure}
\caption{ 
\leftline{ Mean firing rate $f$ dependency of threshold value 
$h_c$ on majority decision mixed state 
$\mbox{\boldmath{$\gamma$}} ^{(s,s/2)}$ }
}
\label{fig:7}
\end{figure}

\begin{figure}
\caption{ 
\leftline{ Mean firing rate $f$ dependency of the threshold value 
$h_c$ on AND mixed state $\mbox{\boldmath{$\gamma$}} ^{(s,s)}$ }
}
\label{fig:8}
\end{figure}

\begin{figure}
\caption{ 
\leftline{ Mean firing rate $f$ dependency of overlap
 $m_{\mu}^{(5,k)}$ as recalled state is mixed state
 $\mbox{\boldmath{$\gamma$}} ^{(5,k)}$ }
}
\label{fig:9}
\end{figure}

\end{document}